**Physics in Perspective**

# The Concept of Fact in German Physics around 1900: A Comparison between Mach and Einstein

Elske de Waal and Sjang L. ten Hagen*

The concept of "fact" has a history. Over the past centuries, physicists have appropriated it in various ways. In this article, we compare Ernst Mach and Albert Einstein's interpretations of the concept. Mach, like most nineteenth-century German physicists, contrasted fact and theory. He understood facts as real and complex combinations of natural events. Theories, in turn, only served to order and communicate facts efficiently. Einstein's concept of fact was incompatible with Mach's, since Einstein believed facts could be theoretical too, just as he ascribed mathematical theorizing a leading role in representing reality. For example, he used the concept of fact to refer to a generally valid result of experience. The differences we disclose between Mach and Einstein were symbolic for broader tensions in the German physics discipline. Furthermore, they underline the historically fluid character of the category of the fact, both within physics and beyond.

*Key words:* Facts; Theory; Ernst Mach; Albert Einstein; Epistemology; Physics.

## Introduction

In recent years, supporters of science have been on the barricades defending the authority of science and its facts. Among the slogans of the 2017 *March for Science* were expressions such as "trust scientific facts, not alternative facts" and "science is not an alternative fact" (figure 1). The message behind these slogans seems straightforward, but is actually complex when we realize that there have been and still are many different ideas about what a "fact" is. Illustratively, the *Oxford English Dictionary* lists ten separate entries for "fact," and there are many subtle differences within these entries. Definitions range from "the sum of circumstances and incidents of a case, looked at apart from their legal bearing" to "that which is known (or firmly believed) to be real or true."[1]

* Elske de Waal is currently finishing a master's degree in history and philosophy of science and about to start a PhD project at Utrecht University. Sjang ten Hagen is a PhD candidate at the University of Amsterdam, specializing in the history of physics and historiography.

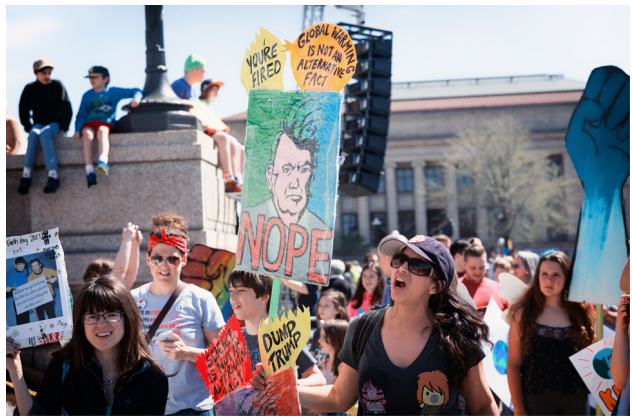

**Fig. 1.** "Science Is Not an Alternative Fact." Photo taken by Lorie Shaull during the Minnesota March for Science on April 22, 2017. *Credit*: Wikimedia Commons

Although various knowledge disciplines claim to rely on facts, great differences exist between the specific meanings of a "fact" in different disciplines. As Lorraine Daston has also noted, a physiological experiment and a sociological questionnaire seem to establish very different things, "but in both cases they are said to provide 'facts.'"[2] Likewise, history and physics are both fact-oriented disciplines, but the facts of history and the facts of physics are of a different kind.[3] Interpretations of the concept of fact do not only vary between disciplines, but also within the very same discipline, or even inside a single research institute. In contemporary particle physics, for example, some understand a "fact" to be an explanation for or generalization of particular experimental phenomena, while others regard "facts" to refer to the phenomena themselves. These contrasting definitions were implied during interviews we conducted with particle physicists based at the Dutch National Institute for Subatomic Physics (Nikhef) in Amsterdam. When asked about the Higgs boson, one of the interviewees explained that it first was a hypothesis, until it was measured, then "it became a fact."* Indeed, even within single disciplines, multiple interpretations of the concept of fact may coexist.

These different interpretations of the fact have a history. Over the past centuries, the concept of fact has been modified repeatedly to serve divergent purposes. Many contemporary features of the fact, such as its association with truth, were not self-evident: it was a seventeenth-century invention to define facts as true. Before then, facts could also be false.[4] Furthermore, researchers only established a link between facts and notions of subjectivity and objectivity over the course of the nineteenth century.[5] So, when the philosopher of science Ludwik

---

* The physicists were Auke Pieter Colijn and Ivo van Vulpen. Transcripts of the interviews (conducted in Amsterdam in January 2018) are available upon request.

Fleck began his influential 1935 book *Genesis and Development of a Scientific Fact* by asking, "What is a fact?," his answer—"A fact is supposed to be distinguished from transient theories as something definite, permanent, and independent of any subjective interpretation by the scientist. It is that which the various scientific disciplines aim at. The critique of the methods used to establish it constitutes the subject matter of epistemology"[6]—would have sounded unfamiliar to scholars and scientists from earlier time periods. Thirteen years before Fleck, Ludwig Wittgenstein posited in his *Tractatus Logico-Philosophicus* (1922) that "the world is the totality of facts, not of things," defining a fact as "what is the case."[7] The differences between Fleck and Wittgenstein's views are subtle, but significant. For one thing, Fleck's facts need to be established—his primary example being "the fact that the so-called Wassermann reaction is related to syphilis"[8]—while those of Wittgenstein are already the case. The major point that we want to convey is that neither Fleck nor Wittgenstein's "fact" was timeless or self-evident.[9]

As it turns out, the best-known literature on facts, including the books by Fleck and Wittgenstein from which we have quoted, has been insensitive to the insight that the concept of fact has been historically fluid, acquiring various interpretations across geographical and disciplinary boundaries. For example, Steven Shapin and Simon Schaffer have famously described early modern practices of fact-making, but without investigating the concept of "fact" itself.[10] The same goes for Bruno Latour and Steve Woolgar's influential book about the construction of scientific facts.[11] Looking back at his earlier work, in 2004, Latour even acknowledged that he had accepted "much too uncritically what matters of fact were." This he blamed on "remaining too faithful to the unfortunate solution inherited from the philosophy of Immanuel Kant."[12]

Since the fact has been subject to many conceptual transformations, various histories of the fact can be told. In recent years, historians of the sciences and humanities have studied its historically fluid meaning in relation to different knowledge practices, including but not limited to philosophy. The focus of historical case studies has ranged from sixteenth-century law to nineteenth-century chemistry, and from seventeenth-century natural history to eighteenth-century statistics.[13] In this article, we build on this historiography of the "fact" by examining how the concept of fact was used in the disciplinary context of German physics around 1900. We focus mainly on the concepts of fact of the physicists Ernst Mach (1838–1916) and Albert Einstein (1879–1955).

Mach and Einstein are no unfamiliar combination. The exact influence of Mach on Einstein, as well as its development during the course of Einstein's career, have long been debated.[14] With the benefit of hindsight, we can identify Mach's historical criticism of the mechanical worldview and the associated concepts of absolute time and space, as most famously outlined in his *The Science of Mechanics* from 1883, as heralding the relativity revolution occurring in the first two decades of the twentieth century. Furthermore, Einstein was deeply influenced by Mach's epistemology, even though he became increasingly skeptical of its merit over the course of his career. In the following, we shed light on the changing relationship between Mach and Einstein's physics and epistemology by comparing their use and understanding of the category fact.

Such a comparison assists the study of the modern notion of a "scientific fact," and how and why it should be distinguished from an "alternative fact." This is because Mach and Einstein's concepts of fact were constitutional for later and current notions, also outside of the physics discipline. Mach's fact-oriented empiricism was a primary source of inspiration for logical positivism and conventionalism, which in turn became hugely influential in shaping twentieth-century philosophical debates about realism, the relation between theory and experiment, and the role and status of scientific facts.[15] Einstein's physics and philosophy, in particular his theory of relativity and his critique of quantum mechanics, also became an essential point of reference in such debates. What is more, Einstein actively contributed to epistemological discussions himself.[16]

Half a century ago, Gerald Holton touched upon the main issue addressed by this paper. In 1968, Holton claimed that there was a "divergence between the conception of 'fact' as understood by Einstein and 'fact' as understood by a true Machist."[17] According to Holton, this divergence related to the status of laws, concepts, and principles, which Mach, unlike Einstein, systematically distinguished from facts. In 1984, Paul Feyerabend criticized Holton's claim; he argued that the contrast between Mach's and Einstein's ideas about the nature of facts should not be exaggerated. On many occasions, Mach used "the term 'facts of experience' in a rather general way," Feyerabend noted.[18] On closer inspection, the evidence presented by Holton and Feyerabend in support of their respective claims proves fragile. Holton focused mostly on Einstein, and admitted that he had only "[re-viewed] briefly the essential points of Mach's philosophy."[19] Feyerabend's response relied on a more in-depth study of Mach's work. However, he projected his own understanding of a fact onto Mach's writings, rather than reconstructing Mach's specific use of the term.[20]

Einstein and Mach's concepts of fact have not been further compared since Feyerabend assessed Holton's claim. Here, we continue the discussion initiated by Holton and Feyerband with the benefit of being able to take into account insights provided by recent scholarship on the history of the fact in German physics.[21] Although neither Mach nor Einstein ever formulated an explicit definition of a fact, unlike Fleck and Wittgenstein, it proves possible to extract coherent interpretations from their writings. To this end, we have taken the original German texts as the basis of our analysis, checking these for the use of the term *Tatsache(n)*. Mach and Einstein used other relevant words as well, such as "*Faktum*" or "*Sachverhalt*," but less frequently.[22]

In what follows, we first discuss how, around 1800, the concept of fact rose to prominence in German intellectual culture, particularly in the discipline of physics. Subsequently, we focus on Mach, analyzing the meaning and role he attributed to facts in relation to his physics. After that, we compare Mach's concept of fact to Einstein's. As it turns out, Mach and Einstein's contrasting understandings of the concept of fact mirrored broader epistemological tensions in the German physics discipline around 1900.

## The Origins of the Fact in German Physics

The concept of fact emerged in sixteenth-century England. It was first used by lawyers, who interpreted a "matter of fact" as a human act or event of which the actual occurrence was in contention. In court, a fact could be decided to be either untrue or probable, depending on the evidence. During the seventeenth century, the concept of fact began to be used in the study of nature as well. In contrast to lawyers, English natural philosophers regarded facts as true in principle.[23] During its subsequent circulation into different languages and disciplines, the meaning of the concept of fact remained fluid.

In the German-speaking context, the concept of fact was widely adopted only around 1800, when the popularity of the then-novel terms *Thatsache* (translated from the English "matter of fact") and *Faktum/Factum* (from the Latin *factum*") rapidly increased.[24] Over the course of the nineteenth century, *Tatsache* (eventually spelled without the "h") became the word most frequently used. The origins of this neologism were in theological history, but it was soon appropriated into the vocabularies of other knowledge domains.[25]

In the late-eighteenth century, prominent German naturalists, including Georg Lichtenberg (1742–1799), adopted the concept of fact to reflect upon the methods of *Naturlehre* (a knowledge domain preceding the modern discipline of "physics" in the German-speaking tradition). Lichtenberg defined facts as real, eternally valid, and strictly empirical. This put them in contrast to theories and hypotheses, which he considered fallible and short-lived.[26] After 1800, the contrast between fact and theory became increasingly common. It was propagated, for example, by Lichtenberg's student Alexander von Humboldt, who argued that "*Thatsachen* are fixed when the fleetingly improvised theoretical building has long collapsed."[27] Lichtenberg and his contemporaries identified facts as the empirical basis of scientific knowledge. The establishment of facts, however, they did not consider a goal of science proper. Lichtenberg held that the gathering of facts should always be followed by synthesis. In his lecture notes, Lichtenberg argued that "it is useless to find facts [*Thatsachen*], when one does not try to bring them in relation to one another. We must certainly have facts, since they are the basis of science [*Wissenschaft*], but they are not science [*Wissenschaft*] itself."[28]

Facts became particularly important in the context of discipline formation at German universities during the beginning of the nineteenth century. The first generation of German "physicists" started to interpret facts by themselves as representing proper scientific knowledge. They emphasized that their novel discipline focused only on facts, which they still defined as real, eternally valid, and strictly empirical. In so doing, they demarcated physics from the speculative enterprise of German idealistic philosophy (*Naturfilosofie*), as represented by Friedrich Schelling and Georg Hegel. Such demarcation strategies were particularly employed by experimentalists such as Gustav Magnus (1802–1870). Like Lichtenberg, Magnus maintained a strict separation between facts, which characterized physics, and theories, which belonged to mathematics.[29] As historian of science Kenneth Caneva has put it, the prevalent ideals in the early nineteenth-century German physics community were "atheoretical." Physicists demanded "facts and experiments without any hypothetical interpretation," and with as little mathematics as possible.[30]

Through practical laboratory training offered by Magnus and others, these fact-oriented, atheoretical ideals became deeply entrenched in the nineteenth-century physical discipline.[31] Not all aspiring physicists, however, could identify with a discipline that was mainly concerned with the establishment of individual, atheoretical facts. In 1871, Hermann von Helmholtz lamented that his former teacher Magnus had only focused on finding new facts, and systematically underappreciated the value of mathematical theory. According to Helmholtz, this was because he had overreacted against the speculative methods of *Naturfilosofie*, which were still very influential when Magnus began to teach at the University of Berlin in the 1830s. Helmholtz recognized the many "new and often remarkable facts" that Magnus had discovered and "brought in connection with the great fabric of science," but found it time to abandon "excessive empiricism which sets out to discover facts which fit to no rule."[32] Helmholtz considered experimental and mathematical physics, as well as facts and theories, to be in a more fluid, continuous relation. In a similar vein, Magnus's former student and laboratory assistant August Kundt pleaded for "an extension of facts in order better to support different fundamental theories or to direct [theory] into new paths" during his inaugural address at the Prussian Academy of Sciences in 1889.[33] Such statements illustrate that, in the late-nineteenth century, empirically minded physicist could focus on facts, and at the same time acknowledge the fruitfulness of (mathematical) theory. This does not imply, however, that the definition of a fact was changing; Helmholtz and Kundt still considered facts to be atheoretical.

Mach and Einstein's conceptions of fact must be understood in relation to this generally perceived tension between fact and theory. Finding an appropriate balance between the two was a pressing issue for both physicists. But as we will show, they adopted divergent definitions of the very concepts of fact and theory. Whereas Mach retained the traditional distinction between fact and theory, Einstein did not.

## Mach's Concept of Fact

As recent research has shown, Mach was a versatile researcher.[34] He was first and foremost a physicist, but had a broad conception of his discipline. Indeed, Mach fashioned himself as a physicist "unconstrained by the conventional barriers of the specialist," as Richard Staley has argued.[35] In his research as well as in his teaching, Mach sought to connect physics to psychology and physiology, and he approached these disciplines within one overarching epistemological framework.[36] Furthermore, Mach believed that contemporary physics was to be practiced from a "historical-critical" perspective, which is why much of his work in physics was interspersed with epistemological reflections.[37] Conversely, Mach's epistemology developed in the context of contemporary debates in physics about mechanics, thermodynamics, and energy conservation.[38]

Mach regarded empirical observation as the primary source of scientific knowledge, and he deemed knowledge acquired by non-experimental means invalid. He found that observations, not speculations, should be the starting point of physics. Still, as recent studies have shown, labels such as "positivism" and "empiricism" do not adequately capture Mach's epistemology.[39] Mach concisely summarized his main epistemological position during a lecture in 1882, claiming that "the goal which [physical science] has set itself is the simplest and most economical expression of facts."[40] To properly understand what Mach meant by this, we need to answer two questions: What are facts, according to Mach? And how does one express them in the most economical way? To answer these questions, we will first give a brief introduction to Mach's epistemological vocabulary. This is crucial, for Mach's understanding of the term fact depended on his specific use of other terms, in particular "sensations" and "elements."

### Elements and Sensations

Mach maintained, as Erik Banks has put it, that "the natural world is made up of individualized events embedded in real causal–functional relations to each other. These events and causal–functional relations are what really exist, and the rest (objects, extended bodies, fields…) are constructed out of them."[41] Mach called these events "elements" (*Elemente*). He considered such elements the fundamental building blocks of experience, and, moreover, of reality. "The very simplest components of what we experience and live through which we do not know how to divide further are called *elements*," he explained.[42] Examples of elements included "colors, tones, pressures, spaces, [and] times."[43] In one of his personal notebooks, Mach further explained: "Colors, space, tones etc. These are the actual realities. There are no others."[44]

The notion of "sensation" (*Empfindung*) was another main ingredient of Mach's epistemological vocabulary. Mach defined sensations as a particular kind of elements, namely as those occurring in the nervous system.[45] It may thus be said that Mach used the terms *elements* and *sensations* to refer to the constituent parts of the outer (physical) and the inner (psychical) world, respectively. But it would be wrong to suppose that Mach considered elements and sensations different in nature. Indeed, for Mach there was no principal distinction between elements and sensations, since he regarded the dividing line between the realms of the physical and the psychical as indiscernible. As he put it himself: "we resolve the whole material world into elements which at the same time are also elements of the psychical world, and, as such, are commonly called sensations."[46] Mach's main reason for sometimes using the term "sensations" (instead of "elements") was to avoid confusion: "we only use the additional terms 'sensations' to describe the elements, because most people are much more familiar with the elements in question as sensations."[47] Despite Mach's attempts to provide clarity, the exact relationship between elements and sensations remained ambiguous. It is certain, however, that Mach considered elements more fundamental than sensations.

Philosophers have paid substantial attention to the meanings of the Machian notions of elements and sensations. Mach's understanding of a "fact" (*Tatsache* or *Faktum*), however, seems to have been mistaken as self-evident. Although Mach's interpretation of a fact closely depended on his understanding of elements and sensations, it would be incorrect to treat Mach's facts as synonymous with either one of these terms, as some have done.[48]

### Facts as Complex Combinations of Elements

According to Mach, facts were *combinations of elements*. He construed facts and elements as equally real.[49] He argued that "all physical facts are made up, ultimately, of the same [*sinnlichen Elementen*]* (colors, pressures, spaces, times)."[50] As elements could be both physical and psychical, facts could also be physical or psychical. Illustrative for this is Mach's assertion that "even the wildest dream is a fact as much as any other."[51]

It seems unlikely that Mach considered the elements themselves to be facts as well. This is because he regarded elements to be indivisible, and facts as "complex." More complex facts could be combinations of simpler facts, which, in turn, could be reduced to elements. For example, the complex facts of sound and water waves can be reduced "to the few facts of wave motion."[52] In Mach's opinion, the most developed sciences "are those whose facts are reducible to a few numerable elements of like nature."[53]

In the 1882 lecture in which Mach claimed that physics entailed "the simplest and most economical expression of facts," he also argued that facts are "exhibited" ("*klargelegt*") by experience.[54] In another lecture, Mach spoke about the revelation ("*Offenbarung*") of facts.[55] These statements give the impression that Mach considered facts to be revealed or discovered, which would mean that, according to Mach, they exist independently from their observation. The statements can also be read differently, however. Mach might also have meant that senses produce facts, and hence that facts are the product of experience. Such a reading is endorsed by Mach's recognition of the senses as the *source* of facts: "We know of only one source of immediate revelation of scientific facts—our senses."[56] But such a reading becomes problematic again when noting that Mach defined observation as the "adaptation of thoughts to facts."[57] This grants facts at least some degree of autonomy, like elements. If we put aside these ambiguities, we can establish that Mach in any case considered facts as inextricably linked to the senses.

---

* In the cited source the phrase "*sinnlichen Elemente*" has been translated, understandably but confusingly, as "sensational elements." "Sensory elements" would have been a less confusing translation.

Due to the complexity of facts, the reproduction ("*Nachbildung*") of a fact in thought during observation is never perfect. Mach frequently stressed that a fact can never be represented in its entirety, and that observation always renders an incomplete representation of a fact. Hence, the exact appearance of a fact in thought heavily depends on the way that it is observed, and also on the observer. In Mach's own words: "*What* aspects of a fact are taken notice of, will consequently depend upon circumstances, or even on the caprice of the observer. Hence there is always opportunity for the discovery of new aspects of the fact."[58]

Let us try to further clarify Mach's understanding of a fact by giving some examples of what he considered facts and what not. First, in an 1891 essay in *The Monist*, Mach compared the "motions of atoms" to the "green of trees." In the latter, he said, "I see a (sensory) fact, in the former a *Gedankending*, a thing of thought."[59] For Mach, the idea of an atom had little scientific value. He considered it a product of speculation, rather than a (representation of a) fact. The green of the trees, on the other hand, he did consider a fact, because it could be directly registered by the senses. Interestingly, Mach left open the possibility that facts have a "nucleus" which lies beyond the reach of experience: "One might be of the opinion, say, with respect to physics, that the portrayal of the sense-given facts is of less importance than the atoms, forces, and laws which form, so to speak, the nucleus of the sense-given facts."[60] Yet he argued that it was not the task of physicists to speculate about this nucleus, or to try to determine the cause or object behind the appearance of a given fact: "unbiased reflexion discloses that every practical and intellectual need is satisfied the moment our thoughts have acquired the power to represent the facts of the senses completely. Such representation, consequently, is the end and aim of physics."[61]

Other examples of facts in Mach's work were "the augmentation of the D-line in the solar spectrum by the interposition of a sodium lamp,"[62] "a real spectrum-image … with Fraunhofer lines occurring in it," and "the charges of conductors residing on the surface."[63] These examples indicate that Mach did not demand facts to be formulated exclusively in terms of elements, as long as they could be reduced to them. Put differently, facts could be interspersed with theoretical tools. In Mach's words, "facts … are extended and enriched, and ultimately again simplified, by conceptual handling."[64]

We now have a grasp of how Mach interpreted the concept of fact and where the ambiguities in his interpretation lay. To Mach, facts were real, complex combinations of elements, and given by the senses (we have noted the ambiguity of this particular feature of a Machian fact). During observation, facts were reproduced in thought, yet only partially due to their complexity. We showed that Mach allowed facts to be supplemented by "conceptual handling." Although the latter might suggest otherwise, Mach did systematically distinguish between fact and theory.[65]

### Theory Orders Facts Economically

Since an individual observer can only experience a limited number of facts, Mach insisted on the importance of communication between individuals. "By communication, the experience of *many* persons individually acquired at first, is collected in *one*," he argued. "The communication of knowledge and the necessity which everyone feels of managing his stock of experience with the least expenditure of thought, compel us to put our knowledge in economical forms."[66] In order for natural knowledge to be shared, facts must thus be presented in an efficient, manageable, or "economical" manner. To this end, Mach attributed an important role to theory.

For Mach, theory meant the adjustment of fact-representing thoughts to each other, in order to find more general representations of facts, for example in the form of natural laws. "To save the labor of instruction and of acquisition," Mach argued, "concise, abridged description is sought. This is really all that natural laws are."[67] Mach defined the ideal theory as "a complete and systematic representation of the facts," but denied that such a theory was ever fully attainable, given the limited accessibility of facts.[68] As science progressed, the ideal was only approached asymptotically. As examples of near-to-ideal theories, Mach mentioned d'Alembert's equations, "which comprise all possible dynamical facts, or the equations of Fourier which comprise all conceivable facts of thermal conduction."[69]

Mach stressed the limitations of theoretical tools, despite their indispensable role of ordering facts economically. However valuable to science, theoretical tools were inevitably contingent and arbitrary, unlike the real elements and facts they aimed to describe. In Mach's opinion, "theories are like dry leaves which fall away when they have long ceased to be the lungs of the tree of science."[70] With regard to laws, he claimed: "in reality, the law always contains less than the fact itself, because it does not reproduce the fact as a whole, but only in that aspect of it which is important to us, the rest being either intentionally or from necessity omitted."[71] About concepts, Mach made similar remarks: "Our physical concepts, however close they come to the facts, must not be regarded as complete and final expression of these facts."[72]

These comments about the limits of theoretical tools illustrate once more that, for Mach, facts were complex and multifaceted. They could be represented only partially, be it by observation or theoretical intervention. In *The Science of*

*Mechanics,* Mach explained this as follows: "A rule, reached by the observation of facts, cannot possibly embrace the *entire* fact, in all its infinite wealth, in all its inexhaustible manifoldness [*Mannigfaltigkeit*]; on the contrary, it can furnish only a rough *outline* of the fact, one-sidedly emphasizing the feature that is of importance for the given technical (or scientific) aim in view."[73] Thus, according to Mach, theories were only capable of expressing a limited aspect of the facts. Following these definitions of fact and theory, Mach regularly emphasized "how important it is to distinguish sharply between concept and law on one side and fact on the other."[74]

Even though Mach made a clear distinction between fact and theory, like most physicists in the nineteenth-century German context, he acknowledged that the two crucially depended on one another. For Mach, facts alone had little scientific value, since the aim of science and physics was the economic ordering of facts. This ordering, in turn, could only be achieved by means of theorizing. Although Mach might not have believed that theories directly represented the real, he did believe they had an essential function. Therefore, Mach's reputation as a positivist only interested in facts is an oversimplification.

The close interdependence of fact and theory in Mach's epistemology becomes further apparent from his understanding of "principles." Mach considered facts and principles as intimately related. In the *Science of Mechanics*, Mach noted that principles were "merely … the ascertainment and establishment of a *fact*."[75] The principle of virtual displacements, for instance, followed instinctively from experience. It was a generalization of the recurring fact that "heavy bodies, of themselves, move only downwards." As Mach explained, "there is contained in the principle of virtual displacements simply the recognition of a fact that was instinctively familiar to us long previously, only that we had not apprehended it so precisely and clearly."[76] In the first chapters of the *Science of Mechanics*, Mach traced the historical origins of some other basic principles of statics, dynamics, and mechanics. He consistently emphasized how principles that seemingly relied on *a priori* arguments had actually been grounded in experience. Mach also argued that once a principle had been established, it could function as a tool in the process of understanding and explaining through economical description, similar to other theoretical tools such as laws and concepts, and unlike the facts ascertained by the principle. In sum, Machian principles played a mediating role between fact and theory: they were closely aligned to experiential facts, while at the same time contributing to their economical description.

In Mach's view, the ideal theoretical construct was a direct expression of facts. Mach believed he had reached this ideal for the concept of mass. In the *Science of Mechanics*, he presented a definition of mass relying on the mutual attraction of bodies. He portrayed this conception of mass as purely factual: "in the concept of mass no theory of any kind whatever is contained, but simply a fact of experience."[77] But even this atheoretical concept remained principally fallible,

Mach conceded: "it is very improbable, but not impossible, that it will be shaken in the future."[78] Generally, Mach argued that any theoretical description, even the most "economical" ones, such as his concept of mass or the mathematically simple laws of d'Alembert and Fourier, "may be deranged by a single newly discovered fact."[79] Of all theoretical tools, Mach seems to have had most trust in the durability of principles, as these corresponded most closely to fact. Yet, even principles were fundamentally imperfect, Mach warned, since the intuitive knowledge that they expressed "is just as fallible as the distinctly conscious."[80]

## Toward a Theoretical Physics

Mach's views on the relation between fact and theory resembled broader opinions in late-nineteenth-century physics. Mach's epistemology, although unique in its specific elaboration, was heavily indebted to how the generation of German-speaking experimental physicists before Mach, including Lichtenberg and Magnus, had defined their emerging discipline. Indeed, Mach conformed to their identification of physics as a non-speculative and fact-oriented discipline. Erik Banks has rightfully pointed out that Mach's "anti-metaphysics (so often associated with Mach himself) was actually the rallying cry of the generation before Mach in their crusade against *Naturphilosophie*."[81]

Furthermore, Mach was a proponent of what John Heilbron has identified as *fin-de-siècle* "descriptionism," which concerned a "withdrawal from big questions and relaxation of claims to knowledge of truth."[82] In Mach's view, as reconstructed by Heilbron, science "had nothing to do with truth, but only with precise, economical and simple description."[83] In a similar vein, Mach's colleague physicist Gustav Kirchhoff (1824–1887) emphasized that mathematical physics was "just a matter of indicating the phenomena which take place, not of determining their causes."[84] Heilbron and others have demonstrated that such views found wide appeal among physicists around 1900, throughout Western Europe.[85]

Descriptionists like Mach and Kirchhoff did not seek underlying causes of observed facts; their primary aim was to represent and order them as efficiently as possible. To this end, they used different techniques. In the case of Mach, descriptionist practices involved experimental work. Mach's photographic experiments on shock waves, for example, were representative for his desire to "lay bare facts" as directly as possible.[86] As Mach explained, the photographic study of shock waves enabled to "exhibit … the individual phases of a movement … to our perception as slowly as we like."[87] A second kind of descriptionist practice characterizing Mach's work concerned the clarification of physical concepts, as in the case of his redefinition of the concept of mass, or of "temperature."[88] A third descriptionist practice—one in which Mach personally was less involved but which was championed by Kirchhoff and Heinrich Hertz—concerned the mathematization of fact-ordering physical concepts, in order to liberate them from metaphysical components.[89] In the preface to the later republication of his 1871 lecture on the history of the principle of the conservation of work, Mach praised Kirchhoff's physics and epistemology, which he regarded as similar to his own. "It was a ray of hope," Mach wrote, "when Kirchhoff pronounced, in 1874, the problem of mechanics to be the complete and simplest description of motions, and this nearly corresponded to the economical representation of facts."[90]

Mach's epistemology thus aligned with those of his contemporaries. Can the same be said of his particular interpretation of the concept of fact? Above we established that early nineteenth-century German physicists made a conceptual distinction between fact and theory; they considered facts as real and lasting, but theories as fallible and short lived. Mach adopted this distinction, and so did the majority of German-speaking physicists in the late-nineteenth century. Hertz, for example, in an 1888 commentary on electric force (which he deemed an obsolete concept), argued that "the facts here communicated are true independently of the theory, and the theory here developed depends for its support more upon the facts than upon the explanations which accompany it."[91]

So, Mach's concept of fact matched broader trends in the disciplinary context of German physics. The details of Mach's use of the term fact, however, were less common; the Machian fact was embedded within a unique conceptual framework, including idiosyncratic notions such as elements and sensations. Furthermore, Mach's emphasis on the complexity of facts was more reminiscent of ideas prevalent in nineteenth-century German historiography than in physics. Like Mach, German historians such as Leopold von Ranke and Wilhelm von Humboldt encouraged the study of historical facts from multiple sides, to represent them as completely as possible, and to extend them in thought. The difference was that historians defined facts as complexes of human rather than natural events (or Machian elements).[92]

Around 1900, the descriptionist views so popular among German experimentalists were losing ground. This was due to the rising status of theoretical physics, a novel subdiscipline that incorporated elements from both experimental and mathematical research traditions.[93] The new physics brought along new epistemologies. Prominent theoretical physicists like Max Planck (1858–1947) proclaimed that the ultimate aim of science was not to represent facts as efficiently as possible, but to find underlying causes and universally valid laws, if necessary by using speculative methods and the power of imagination.[94] This placed Planck directly opposite Mach and his conception of proper scientific method, and more in agreement with Mach's contemporary Helmholtz.[95] For Mach, the real was confined to actual "elements" and combinations of elements ("facts"). Planck, on the other hand, defined the real in terms of a "world-picture" (*Weltbild*), which was not to be found in Machian facts or elements, but only in universally valid laws and principles.

In 1908, Planck openly contested the still-influential views of his Austrian colleague. In a lecture entitled *The Unity of the Physical World-Picture*, which was a direct attack on Mach and his epistemology, Planck asked rhetorically: "Is there today a single physicist worthy of serious consideration who doubts the reality of the energy principle? Rather, the recognition of this reality is nowadays a prerequisite for winning any scientific respect." About Mach's philosophy, he remarked that it "does not affect the essence of natural science. This is because the outstanding characteristic of all scientific research—the demand for a *constant* world picture, independent of changing times and peoples—is alien to it."[96] Indeed, Mach had at one point argued that the energy principle "consists in a special form of viewing facts, but its domain of application is not unlimited," thereby questioning its universality.[97]

Initially, Einstein chose Mach's side, for example by calling Planck's criticism "unjust" in 1913.[98] In subsequent years, however, he came to embrace many of Planck's points of criticism.[99] Like Planck and other theoretical physicists, Einstein became increasingly convinced that mathematically invented theories could directly represent reality, a notion that was reinforced by the completion of general relativity in 1915.[100] In the following section, we show that Einstein's rapprochement to Planck's views and his growing aversion to Machian epistemology were accompanied by the development of a fundamentally different understanding of the term fact.

## Einstein's Concept of Fact

Initially, Einstein complied with the epistemologies of German experimentalists like Lichtenberg, Magnus, and Mach. He distinguished between "observable fact" ("*beobachtbare Tatsache*") and the "merely conceptual" ("*bloss begriffliches*").[101] The young Einstein would generally stress that any theory had to correspond with the facts while adding to that, in Machian fashion, that such correspondence would never be perfect: "Of course, an exact agreement with the facts is out of the question," he noted in 1906, while discussing recent developments in molecular theory.[102] In 1914, during his inaugural lecture at the Prussian Academy of Sciences, Einstein emphasized the importance of principles in physics. He did so while defining these principles as inductions from "complexes of facts" (*Komplexen von Erfahrungsthatsachen*).[103] In another lecture at the Prussian Academy in 1921, entitled *Geometry and Experience*, Einstein expressed the opinion that mathematical theories were never direct representations reality: "As far as the propositions of mathematics refer to reality, they are not certain; and as far as they are certain, they do not refer to reality."[104]

All this seems perfectly compatible with Mach's economical approach to physics. Yet, Einstein's alignment with Mach proved inconsistent. In contrast to Mach and like Planck, Einstein attached much value to the idea that physical theories and principles could be "real," even early in his career. In the 1914 inaugural lecture, for example, he claimed that it was possible that "theoretical principles correspond to reality."[105] Furthermore, Einstein maintained that theories and principles could also be rationally invented and then tested experimentally, instead of the other, inductive way around; he argued that theories of gravitation should be confronted with experimental facts only after they had been mathematically invented.[106]

Over the course of his career, Einstein embraced the potential of mathematical theory to fully grasp reality.[107] Important backgrounds of Einstein's shift toward this position were his campaign against quantum mechanics and his pursuit of a unified field theory. Such a mathematically simple theory he depicted as an unavoidable alternative to the messy, experiment-driven, and probabilistic theory of quantum mechanics.[108] The more deeply Einstein became involved in his unification project and the more heavily he criticized quantum mechanics, the less he came to appreciate the heuristic role of experiment. Indeed, Einstein regarded quantum theory to be superficial and uninspired precisely because it was designed to fit experimental outcomes.[109] In 1925, furthermore, Einstein wrote to his friend Paul Ehrenfest that "inductive means will never get you to a sensible theory."[110] In 1933, he insisted that "any attempt logically to derive the basic concepts and laws of mechanics from the ultimate data of experience is doomed to failure."[111] Some years later, in 1936, Einstein blamed nineteenth-century physicists, including Mach, for their "failure to understand" that "there is no inductive method which could lead to the fundamental concepts of physics."[112]

## Facts and Principles

Clearly, Einstein abandoned his faith in Mach's economical, fact-ordering approach to physics later in his career. Meanwhile, Einstein had also appropriated an understanding of the concept of fact fundamentally different from that of his Austrian colleague. This is most evident from Einstein's synonymous treatment of facts and principles. For example, in 1916, while introducing his theory of relativity to a wider audience, Einstein referred to "the fact of the equality of inertial and gravitational mass, which is strongly confirmed empirically."[113]

In a 1918 letter to his friend Michele Besso, "a loyal Machist" who had introduced Einstein to Mach's work,[114] Einstein explicitly blurred the distinction between fact and theory (including principles) that had been propagated by Mach. Einstein seemed upset that Besso, one of his dearest friends, had accused him of developing a speculative theory that was not directly connected to facts. He explained why he found that the theory of relativity was not based on speculation, but on facts. In doing so, he broadened the scope of what he considered factual, ostensibly without being aware of it. "On rereading your last letter," Einstein wrote to Besso, "I discovered something that downright annoys me: speculation allegedly had revealed itself to be superior to empiricism." He continued: "In this regard you are thinking of the development of the theory of relativity. But I find that this development teaches something different that is almost the opposite, namely, that in order to be reliable, a theory must be built upon generalizable *facts*." Einstein then listed some "old examples" of theories and the facts that they had been built upon: the "main theorems of thermodynamics [are based] on the impossibility of the *perpetuum mobile*. Mechanics on the empirically explored law of inertia. Kin[etic] gas theory on the equivalency of heat and mech[anical] energy (also historically)." The theory of relativity, according to Einstein, had been derived from the facts of "the constancy of the velocity of light [and] Maxwell's equations for the vacuum, which on their part are based on empirical foundations. Relativity with regard to unif[orm] translation is an *observed fact*. General relativity: *Equivalency of inertial and gravitational mass*."[115] In 1921, likewise, Einstein called "the constancy of speed of light" and "the equivalence of inertial and ponderable mass" "very definite facts."[116]

The above examples indicate that Einstein was developing a different, broader concept of fact than Mach and his contemporaries. By the beginning of the 1920s, Einstein seemed to consider facts and principles interchangeable, interpreting both as general results of experience. In 1919, Einstein labelled the impossibility of the perpetuum mobile neither a fact nor a principle, but an "*allgemeine Erfahrungsresultat*."[117] Strikingly, the first two examples listed by Einstein in his letter to Besso also figured prominently in Mach's work. In the 1871 lecture, for example, Mach lengthily discussed the impossibility of perpetual motion. Like Einstein, Mach emphasized that the impossibility of constructing a perpetuum mobile had not been derived from *a priori* speculation, but from experience. In contrast to Einstein, however, Mach consistently labelled it as a principle ("*Satz*" or "*Princip*") rather than a fact.[118] As we have argued above, for Mach principles and facts were, though closely related, fundamentally distinct. Einstein's understanding of the concept of fact had thus become more inclusive than Mach's, and included the generalized descriptions of experience that Mach would have referred to as principles. Into what exactly had it developed?

### Three Interpretations of Fact

We have found that Einstein used the term fact in at least three different ways. None of these exactly mirrored the conceptions of Mach and his late-nineteenth-century peers. First, Einstein employed the term fact, usually in the singular form, when referring to a principle based on experience that he considered a universally valid representation of reality, as in his 1918 letter to Besso. But on many other occasions as well, we find examples such as "the fact that the gravitational and inertial masses are identical," or "the fact that weak disturbing forces are able to produce alterations of any magnitude in the physical condition of a system."[119] In 1949, still, Einstein wrote about "the fact of the equality of inert and heavy mass, i.e., the fact of the independence of the gravitational acceleration of the nature of the falling substance."[120] As we argued above, this was at odds with conceptions of Mach and his fellow descriptionists, who never assumed principles, let alone other theoretical tools, to be universally valid. So, while Mach and Einstein could both use the term fact to refer to something they considered to directly represent reality, they had different ideas about what met this demand. As evidenced by the polemic between Mach and Planck, this was not so much a typical difference between Mach and Einstein, but reflected a general tension in German physics around 1900.

Second, Einstein referred to facts, in the plural, when denoting a set of individual sense experiences to which a certain theory needed to correspond.[121] While this did conform to previously prevalent interpretations, Einstein ascribed a less primary role to these kinds of facts than did his nineteenth-century predecessors. This was because he thought experience was not a good starting point for developing a theory that would correspond to reality. From the 1930s onwards, moreover, when referring to individual sense experiences, Einstein generally preferred to use phrases such as "*Erlebnismaterial*," "*Sinnen-Erlebnisse*," or the "*erfahrungsmässig Gegebenen*," rather than "*Tatsachen*."[122]

Third, Einstein used the term "fact" to refer to something beyond discussion. This could be a physical concept or experimental outcome, but also a general statement, such as: "It is a fact [*Tatsache*] that Mach has had tremendous impact upon our generation of natural scientists."[123] In Einstein's work—especially in English translations, but also in the original German—one finds many passages in which the term "fact" ("*Tatsache*," "*Faktum*," or "*Sachverhalt*") is not used reflectively and in relation to physics. Although English translations of his work might give a different impression, Mach rarely used the term in such an unreflective way.[124] Apart from a few exceptions, Mach systematically utilized the term fact (*Thatsache*) to reflect upon the methods of his discipline, just like other nineteenth-century physicists. Einstein's threefold interpretation of the term fact indicates that he applied the concept not just alternatively, but also less systematically than did Mach.

## Conclusion

Despite his growing aversion to Mach's epistemology, Mach's work taught Einstein many lessons. To name one, Einstein recognized Mach's insight that even the most fundamental concepts of physics were historical in nature, which implied that they were not unchangeable. In his 1916 eulogy for Mach, Einstein phrased this insight as follows: "Concepts that have proven useful in ordering things can easily attain an authority over us such that we forget their worldly origin and take them as immutably given.… It is not at all idle play when we are trained to analyze the entrenched concepts, and point out the circumstances that promoted their justification and usefulness and how they evolved from the experience at hand."[125] Einstein had in mind foundational concepts such as inertia, space, and time, which had long been regarded as unalterable foundations of the mechanical world view that Mach had criticized. He may also have thought about physical concepts of heat and energy, the histories of which Mach had studied extensively. Einstein most likely did not have in mind the concept of fact.

Yet, the Machian lesson about the historicity of concepts also applies to epistemological ones, including the fact. In this study we have aimed to demonstrate that Einstein and Mach used and interpreted this concept differently. The difference became increasingly pronounced over the course of Einstein's career, as he developed an increasing aversion to inductive empiricism, which he associated with Mach's economical approach to physics. As Einstein and other theoretical physicists gained confidence in the potential of mathematical theory to correspond directly to reality, his understanding of the concept of fact shifted: it came to comprise the domain of the theoretical as well as the empirical. By dissolving the long-standing distinction between fact and theory, Einstein placed himself outside of the disciplinary tradition of nineteenth-century physics that Mach's epistemology had been an expression of. The contrast we have found between Einstein's and Mach's interpretations indicates that the concept of fact continued to transform after it had claimed its place in the German physics discipline. Given Mach's lessons on the historicity of scientific concepts and the tremendous changes in the foundations of physics in this period, this cannot come as a surprise.

To be able to determine to what extent this contrast between Mach and Einstein really resembled general trends, we encourage further study into the history of the fact, both within and beyond the boundaries of the German physics discipline. Clues for such further study can be found in the observations of a well-known follower of Mach, the logical positivist philosopher Rudolf Carnap (1891–1970). In 1966, Carnap proposed to distinguish more carefully between "facts," which he defined as "particular events," and "laws," which he defined as "universal statements." Implicitly referring to the historicity of these notions, Carnap observed that their meaning had become blurred in recent scientific discourse:

> Scientists often refer to universal statements—or rather to what is expressed by such statements—as "facts." They forget that the word "fact" was originally applied (and we shall apply it exclusively in this sense) to singular, particular occurrences. If a scientist is asked about the law of thermal expansion, he may say: "Oh, thermal expansion. That is one of the familiar, basic facts of physics." In a similar way, he may speak of the fact that heat is generated by an electric current, the fact that magnetism is produced by electricity, and so on. These are sometimes considered familiar "facts" of physics. To avoid misunderstandings, we prefer not to call such statements "facts." Facts are particular events. "This morning in the laboratory, I sent an electric current through a wire coil with an iron body inside it, and I found that the iron body became magnetic." That is a fact unless, of course, I deceived myself in some way. However, if I was sober, if it was not too foggy in the room, and if no one has tinkered secretly with the apparatus to play a joke on me, then I may state as a factual observation that this morning the sequence of events occurred.

When we use the word "fact," we will mean it in the singular sense in order to distinguish it clearly from universal statements. Such universal statements will be called "laws" even when they are as elementary as the law of thermal expansion.[126]

We have quoted Carnap so extensively since his observations directly touch upon the historical transformation of the fact that we have aimed to capture in this study. Most probably, Mach would have agreed with Carnap's call to distinguish more carefully between fact and law. Einstein, on the other hand, conflated these notions, exactly in the way that Carnap described. Moreover, Carnap's analysis suggests that the conceptual transformation from facts as singular events bound to a certain time and place, as in the case of Mach, to facts as universal statements, as in the case of Einstein, resembled broader trends in early twentieth-century science.

We began this article by noting that some consider the inevitability of scientific facts a precondition of scientific authority (as in figure 1). Histories of the fact, including this one, have questioned that premise, by showing that there never was a single preferred interpretation of a fact; multiple concepts of fact have developed and co-existed over the past centuries. To those considering the meaning of a fact self-evident, this may be a distressing message. But it may also create opportunities. We think it is important that those relying on the authority of scientific facts reflect on what facts can be and what not, instead of assuming, unduly, that this is self-evident. What does a scientific fact mean today? Are facts complex, as Mach assumed, or simple? Do facts represent singular events or universal statements? Are they empirical or theoretical, or can they perhaps be both? Regardless of the different answers given to these questions in different disciplinary contexts, we maintain that reflection on and openness about current notions of fact is vital in confronting the present-day, anti-scientific politics of alternative facts. It creates the opportunity to make clear what distinguishes scientific facts from alternative facts.

## Acknowledgements

We thank Jeroen van Dongen and Daan Wegener for their engagement in the research leading to the present article. We also thank them, as well as Friedrich Stadler, Jaco de Swart, and Jos Uffink for comments on earlier drafts. Finally, we are grateful to the editors of *Physics and Perspective*, in particular to Joseph Martin and Richard Staley, for their support in preparing this article for publication.

third party material in this article are included in the article's Creative Commons licence, unless indicated otherwise in a credit line to the material. If material is not included in the article's Creative Commons licence and your intended use is not permitted by statutory regulation or exceeds the permitted use, you will need to obtain permission directly from the copyright holder. To view a copy of this licence, visit http://creativecommons.org/licenses/by/4.0/.

**Publisher's Note**   Springer Nature remains neutral with regard to jurisdictional claims in published maps and institutional affiliations.

Institute for Theoretical Physics,
Vossius Center for the History of the Humanities and Sciences,
University of Amsterdam
Science Park 904
Postbus 94216, 1090 GE Amsterdam,
The Netherlands
e-mail: sjangleonard@gmail.com